# Large Enhancement of Properties in Strained Lead-free Multiferroic Solid Solutions with Strong Deviation from Vegard's Law

Tao Wang[1,12], Mingjie Zou[2,12], Dehe Zhang[3], Yu-Chieh Ku[4], Yawen Zheng[1], Shen Pan[1], Zhongqi Ren[1], Zedong Xu[5], Haoliang Huang[6], Wei Luo[7], Yunlong Tang[2,8], Lang Chen[5], Cheng-En Liu[4], Chun-Fu Chang[9], Sujit Das[10], Laurent Bellaiche[7], Yurong Yang[3,*], Xiuliang Ma[2,*], Chang-Yang Kuo[4,11,*], Xingjun Liu[1,*] and Zuhuang Chen[1,13*]

[1] State Key Laboratory of Advanced Welding and Joining of Materials and Structures, School of Materials Science and Engineering, Harbin Institute of Technology, Shenzhen, 518055, China

[2] Shenyang National Laboratory for Materials Science, Institute of Metal Research, Chinese Academy of Sciences, Wenhua Road 72, Shenyang 110016, China

[3] National Laboratory of Solid State Microstructures and Collaborative Innovation Center of Advanced Microstructures, Department of Materials Science and Engineering, Nanjing University, Nanjing 210093, China

[4] Department of Electrophysics, National Yang Ming Chiao Tung University, Hsinchu 30010, Taiwan

[5] Department of Physics, Southern University of Science and Technology, Shenzhen 518055, China

[6] Quantum Science Center of Guangdong-HongKong-Macao Greater Bay Area (Guangdong), Shenzhen 518045, China

[7] Physics Department and Institute for Nanoscience and Engineering, University of Arkansas, Fayetteville, Arkansas 72701, USA

[8] School of Materials Science and Engineering, University of Science and Technology of China, Wenhua Road 72, Shenyang 110016, PR China

[9] Max-Planck Institute for Chemical Physics of Solids, Nöthnitzer Str. 40, 01187 Dresden, Germany

[10] Materials Research Centre, Indian Institute of Science, Bangalore 560012, India

[11] National Synchrotron Radiation Research Center, 101 Hsin-Ann Road, Hsinchu 30076, Taiwan

[12] These authors contributed equally to this work.

[13] Lead contact

*Correspondence: zuhuang@hit.edu.cn; changyangkuo@nycu.edu.tw; xlma@imr.ac.cn; yangyr@nju.edu.cn; xjliu@hit.edu.cn;

**SUMMARY**
Efforts to combine the advantages of multiple systems to enhance functionlities through solid-solution design present a great challenge due to the constraint imposed by the classical Vegard's law. Here, we successfully navigate this trade-off by leveraging the synergistic effect of chemical doping and strain engineering in solid-solution system of $(1-x)$BiFeO$_3$-$x$BaTiO$_3$. Unlike bulks, a significant deviation from the Vegard's law accompanying with enhanced multiferroism is observed in the strained solid-solution epitaxial films, where we achieve a pronounced tetragonality (~1.1), enhanced saturated magnetization (~12 emu/cc), substantial polarization (~107 μC/cm$^2$), high ferroelectric Curie temperature (~880 °C), all while maintaining impressively low leakage current. These characteristics surpass the properties of their parent BiFeO$_3$ and BaTiO$_3$ films. Moreover, the superior ferroelectricity has never been reported in corresponding bulks (*e.g.*, $P$ ~5 μC/cm$^2$ and $T_C$ ~300 °C for bulk with $x = 0.5$). These findings underscore the potential of strained $(1-x)$BiFeO$_3$-$x$BaTiO$_3$ films as lead-free, room-temperature multiferroics.

**PROGRESS AND POTENTIAL**
The appeal of multiferroic BiFeO$_3$ lies in its promise for low-power nanoelectronics, yet the road to their technological utilization is greatly hindered by large leakage currents. While doping with insulating BaTiO$_3$ can effectively reduce the leakage response of BiFeO$_3$, it entails significant sacrifices in its ferroelectric characteristic. Our study leverages the synergistic effect of chemical doping and strain engineering to achieve a remarkable breakthrough in the multiferroic properties of $(1-x)$BiFeO$_3$-$x$BaTiO$_3$ films—a feat not realized in the corresponding bulks. More importantly, we have surmounted a significant hurdle by achieving enhanced properties akin to the synergy of two elements creating more than their sum (i.e., 1+1 > 2). In essence, the enhanced properties of the strain solid-solution films surpass those of their parent constituents. These achievements not only push the boundaries of material science but also open new avenues for advancing multiferroic technologies.

# INTRODUCTION

Multiferroic materials have received great attention due to their potential applications in multifunctional nanoelectronics, such as magnetoelectric devices, tunable microwave devices, sensors[1,2]. However, most single-phase multiferroic materials grapple with formidable obstacles including large intrinsic leakage currents and low ordering temperatures, thereby curtailing their potential applications. As one of the few promising single-phase multiferroic materials, $BiFeO_3$ (BFO) possess both ferroelectric ($T_C$ ~830 °C) and antiferromagnetic ($T_N$ ~367 °C) ordering above room temperature[1]. However, similar to other single-phase multiferroic materials, BFO is plagued by high leakage current density that can significantly deteriorate its ferroelectric and magnetoelectric performances, such as decreasing electrical breakdown strength, increasing energy dissipation and weakening device reliability, etc[2]. Theses greatly restrict its technological applications, such as low-power nonvolatile memory and logic devices. To overcome severe leakage issues, BFO often form solid solutions with other more insulating compounds, such as $(1-x)BiFeO_3-xBaTiO_3$ (BFO-BTO), $(1-x)BiFeO_3-xSrTiO_3$, and $(1-x)BiFeO_3-xNa_{0.5}Bi_{0.5}TiO_3$ [3-7]. The properties of these continuous solid solutions are typically evaluated using empirical Vegard's law [8,9]. According to the century-old Vegard's law, the structure of a solid solution can be determined by a linear interpolation of the lattice constants of its constituents, weighted by their respective concentrations (Figure 1A). Despite observing a slight deviation (on the order of a few percent) from linearity [10,11], the overall intrinsic properties of these solid solutions, such as lattice constant, unit-cell volume, spontaneous polarization $P_S$, and $T_C$, hardly surpass those of their two end-member compounds simultaneously. Taking BFO-BTO system as an example, pseudocubic lattice parameter and unit-cell volume of bulk BFO-BTO generally obey the Vegard's law[4,12]. Similar to other continuous solid solutions, their intrinsic physical properties, such as $P_S$ and $T_C$, vary between BFO and BTO, and decrease with the increasing concentration of BTO[4,13]. All told, although bulk BFO-BTO demonstrates reduced leakage compared to BFO, it comes at the expense of robust ferroelectricity inherent to BFO. Therefore, if the constraints imposed by Vegard's law can be surpassed, there exists the potential to enhance comprehensive properties through solid solutions, effectively combining the advantages of multiple systems (i.e., 1+1 > 2).

Here, we report the observation of strong deviation from the Vegard's law in lead-free, multiferroic solid-solution system BFO-BTO *via* synergistic effect of chemical doping and strain engineering (Figure 1). The strained BFO-BTO epitaxial thin films exhibit a pronounced deviation from Vegard's law, showcasing notable enhancements in the out-of-plane *c* lattice parameter, unit-cell volume, and multiferroic properties, all while maintaining low leakage response. These solid-solution films exhibit an ultralow leakage current density, which is four orders of magnitude lower than those observed in the pure BFO film. Of particular significance is our achievement in optimizing the multiferroic attributes without compromising the $P_S$ and $T_C$ of BFO. Importantly, these strained solid-solution thin films display a bolstered polarization ($P_S$ ~ 107 μC/cm²), an elevated $T_C$ (~ 880 °C) and an enhanced magnetism (~ 12 emu/cc); all are higher than the two end-member compounds. The enhanced ferroelectricity in the solid-solution films is attributed to enhanced lone pair state and the increased hybridization between Ti/Fe and O ions in the films, compared to BFO and BTO films. Our endeavor spotlights the immense potential harbored within BFO-BTO films—a groundbreaking step towards lead-free, room-temperature multiferroic materials, which is also illuminating for future studies of potential synergistic effects via the combination of chemical doping and strain engineering.

## RESULTS

**Large tetragonality and strong deviation from Vegard's law induced by epitaxial strain**

~ 60-nm-thick (1-$x$)BFO-$x$BTO ($x$ = 0.0-1.0)/25 nm SrRuO$_3$ (SRO) heterostructures were epitaxially grown on (001) SrTiO$_3$ (STO) single crystal substrates (Experimental section). X-ray diffraction patterns reveal that all films are single phase without any impurity (Figure 1B and Figure S1). The excellent crystalline quality of the BFO-BTO films was demonstrated by the Laue oscillations (inset of Figure 1B) and narrow full width at half-maximum of the rocking curves around the 002-reflection (Figure 1C). In addition, a typical AFM topographic image reveals a smooth surface with a root-mean-square roughness of only ~300 pm (Figure 1D). Interestingly, the solid-solution films exhibit much larger out-of-plane $c$ lattice parameters than the individual BFO and BTO films, particularly for $x$ = 0.2-0.7. To unveil the lattice structures of the films, X-ray reciprocal space mapping studies were carried out around the 103-diffraction conditions for both the films and substrates (Figure 1E and Figure S2). Interestingly, the $c$ lattice parameters (ranging from 4.29 Å-4.34 Å) in these coherently strained solid-solution films with $x$ = 0.2-0.7 are significantly larger, by up to 8.67%, compared to those in the corresponding BFO-BTO bulks (~ 3.98 Å-4.00 Å) [4,12]. Moreover, the lattice structures of these highly-strained solid-solution films appear to be tetragonal, which is confirmed by the single 103-diffraction spot, no half-integer reflection (Figure S3) and same {103} diffraction spot position with varying φ values[14] (Figure S4). It is important to note that, unlike bulks [4,12], the $c$ lattice parameter and unit-cell volume of the films exhibit a strong deviation from Vegard's law (Figure 1F). Particularly, under substrate-induced compressive strain (~ -2.3%), the $c$ lattice parameter is as large as 4.34 Å for $x$ = 0.5, resulting in a $c/a$ ratio of 1.11. Interestingly, under similar strain condition, BFO film exhibits a much smaller $c/a$ [15], while BTO film would experience strain relaxation at a low critical thickness of < 5 nm[16]. These findings underscore the advances of synergistic effect of chemical and strain engineering in manipulating the structure of BFO-BTO films.

**Microscopic structure characterizations using scanning transmission electron microscopy**

To investigate the microscopic-atomic structrue, high-resolution scanning transmission electron microscopy (STEM) analyses were carried out on films with $x$ = 0.3 and $x$ = 0.7 (Figure 2 and Figure S5). The atomic-resolved electron energy loss spectroscopy (EELS) element mapping reveals that Ba/Bi and Fe/Ti atoms are uniformly distributed at the A-site and B-site of the perovskite structure, respectively, implying the formation of a near-ideal solid solution (Figure 2A and E). In addition, the EELS-STEM result shows an atomically sharp interface between the SRO bottom electrode and the BFO-BTO films; further indicative of high film quality. Although the composition distribution of the two kinds of films is uniform, the ratios of Bi/Ba and Fe/Ti within the two films are different, which may lead to different local lattice and polarization states. To verify this, integrated differential phase contrast (iDPC)-STEM technique was employed [17,18]. By using 2D gauss fitting, local tetragonality can be extracted, yielding similar value as depicted in Figure 2B and 2F for $x$ = 0.3 and $x$ = 0.7, respectively. The results are well consistent with the above XRD results that both films have similar $c$ lattice parameter of 4.31 Å and $c/a$ ratio of 1.10. However, completely distinct polarization configurations were observed, as illustrated in Figure 2C-D and 2G-J. For $x$ = 0.3, the polarization is predominantly along the out-of-plane direction, implying a single-domain charactersitic. Such a uniform polar structure is completely different from that observed in the bulk with nanodomains[19], suggesting that strain would effectively modify the domain configuration. In contrast, for $x$ = 0.7, the film exhibits a nanodomain structure with spatially nonuniform polar states, where domains with in-plane polarization components are clearly visible. In addition, the film with $x$ = 0.7 exhibits a smaller polar displacement compared to the $x$ = 0.3 counterpart (Figure 2D and Figure 2H-J). Therefore, these results indicate that although these two films present a nearly identical external lattice structures (*i.e.,* lattice parameters and $c/a$ ratio), their local polarization characteristics exhibit significant differences.

**Enhanced multiferroic properties with ultralow leakage response**

Here we go one step further to analyze the physical properties of the films (Figure 3). Our BFO thin film exhibits a high leakage current density as typically observed in previous studies [20]. After incorporating BTO into BFO, the electrical resistivity of the solid-solution films is greatly improved, where leakage current density of the solid-solution films is nearly four orders of magnitude lower than that of pure BFO (Figure 3A). To further understand the enhanced resistivity in BFO-BTO thin films, we analyzed the bandgap $E_g$ using optical transmission absorption spectrum (Figure S6-7)[21]. Our analysis reveals that the $E_g$ for both pure BFO and BTO thin films are approximately 2.6 eV and 3.18 eV, respectively, which aligns with previous reports[22][21]. When the BTO is incorporated into BFO, $E_g$ of the solid solution films gradually increases to ~3.0 eV. Interestingly, despite the fact that these solid solution films have $E_g$ smaller than that of pure BTO film, they exhibit significantly lower leakage current compared to the pure BTO. This may be attributed to the presence of local compositional disorder within the solid solution, which enhances the local carrier scattering, as observed in high-entropy multicomponent oxides[23]. Consequently, these factors contribute to the ultralow leakage response in the solid solution films.

The presence of high leakage current density in the pure BFO film causes the measurement of the *P-E* loop to appear elliptical in shape (Figure S8A). In contrast, the low leakage response of the solid-solution thin films enable the acquisition of saturated and square *P-E* loops, as depicted in Figure 3B. Notably, for solid-solution films with $x$ ranging from 0.2 to 0.5, $P_S$ reaches up to ~ 107 µC/cm² (Figure 3C). It is a significant improvement compared to bulk BFO-BTO, where $P_S$ generally remains below approximately 40 µC/cm² at $x = 0.2$-0.4 and decreases further for $x > 0.4$ (*e.g.* $P_S$ is ~5 µC/cm² at $x = 0.5$) [4,24]. The $P_S$ of those strained solid-solution BFO-BTO films also surpass those of pure BFO and BTO [Noting that $P_S$ are ~26 µC/cm² for pure BTO thin films and ~60 µC/cm² for BFO thick films grown on (001) STO (Figure S8B)]. Interestingly, despite BFO-BTO films exhibit same *c/a* for $x = 0.3$ and $x = 0.7$, significant different ferroelectric polarizations are observed. Specifically, $P_s$ is ~107 µC/cm² for $x = 0.3$ while it decreases to ~70 µC/cm² for $x = 0.7$, which is in good agreement with the aforementioned STEM results. Additionally, we evaluated $T_C$ of the BFO-BTO thin films by measuring temperature dependence of the *c* lattice parameter (Figure 3D). As demostrated, the $T_C$ of the solid solution films are all above 600 °C, all are much higher than the corresponding bulk values[13]. Remarkably, the $T_C$ of $x = 0.3$ film is determined to be as high as ~880 °C, which is higher than those of pure BFO and pure BTO thin films, suggesting the strong deviation from the Vegard's law. Besides, unlike other compositions, the ferroelectric-to-paraelectric phase transition in the $x = 0.7$ film occurs over a wide temperature range, exhibiting the typical diffuse characteristic for relaxor ferroelectrics[25]. Bulk BFO-BTO with $x \sim 0.33$-0.8 exhibits a typical relaxor behavior, which enhances with incearesing $x$[4,13]. Although compressive strain can suppress relaxor behavior[26], it would be more challenging to fully suppress the relaxor character in BFO-BTO films with $x = 0.7$ than in the films with $x = 0.3$ under similar compressive strain. Consequently, the film with $x = 0.7$ still exhibits relaxor-like behavior, with a nonuniform nanodomain structure (Figure 2G) and a diffuse structrual transition.

We further analyze the magnetic properties of BFO-BTO films at room temperature (Figure 3E). The BFO film exhibits a weak ferromagnetism due to its canted antiferromagnetic order[27]. Interestingly, the films with $x = 0.3$-0.5 exhibit enhanced saturated magnetization compared to pure BFO. It is worth to note that the saturated magnetization of the film with $x = 0.3$ is three times larger that of the BFO film. Similar enhanced magnetization has also been observed in BFO-BTO bulks[27]. In BFO, it is G-type antiferromagnetic due to the half-filled $d^5$ electron configuration, which induces a strong antiferromagnetic superexchange interaction between neighboring $Fe^{3+}$ [28,29]. The increased Fe-O distance along the *c*-direction, compared with BFO, would lead to a decrease in the hopping integral. Additionally, as demonstrated in bulks[27], the incorporation of nonmagnetic BTO into BFO results in the partial replacement of $Fe^{3+}$ sites by $Ti^{4+}$, a process that weakens the antiferromagnetic interaction[27,30] and enhances the saturated magnetization through spin canting in the solid-solution films.

To clearly showcase the enhanced multiferroic properties in strained BFO-BTO solid-solution films, we compare their $P_S$ and $T_C$ with other representative lead-free perovskite multiferroics and ferroelectrics (Figure 3F). The BFO-BTO films exhibit significant advantages in terms of high $P_S$ and $T_C$ compared to others. The ferroelectric performances observed in BFO-BTO films are not only much larger than that in BFO-BTO bulks, but also exceed those of individual BFO and BTO thin films. These excellent ferroelectric properties (*i.e.*, large $P_S$, high $T_C$ and ultralow leakage current density) further highlight the advances of strained BFO-BTO solid-solution thin films for technological applications.

**Linearly polarized X-ray absorption spectroscopy studies**

To gain deep insights into microscopic origins of the large enhancement of ferroelectricity in the solid-solution films, we conducted linearly polarized X-ray absorption spectroscopy (XAS) at the Fe and Ti $L_{3,2}$ edges (Figure 4 and Figure S9). The overall line shapes of Fe and Ti $L_{3,2}$ edges show that Fe and Ti cations are trivalent and tetravalent[30,31], respectively. It is clear that all BFO-BTO films exhibit the linear dichroism (Figure 4A and 4C). The crystal/ligand field divides both $L_3$ and $L_2$ edge into two distinct peaks, namely $t_{2g}$ ($d_{xy}$, $d_{xz}$, $d_{yz}$) and $e_g$ ($d_{3z^2-r^2}$, $d_{x^2-y^2}$). Here we define the energy shift of $\Delta t_{2g}$ and $\Delta e_g$ as the $L_3$-edge peak energy of $E\perp c$ for $t_{2g}$ and $e_g$ minus that of $E//c$, respectively. According to XRD measurements, the pure BTO film experiences a compressive strain with $c/a$ ratio of ~1.02. The strain-induced elongation of the local TiO$_6$ clusters along the $c$-axis (*i.e.*, $z$-axis) would result in the orbital energy levels of $d_{xz}/d_{yz} < d_{xy} < d_{3z^2-r^2} < d_{x^2-y^2}$ (*i.e.* $\Delta t_{2g} > 0$ and $\Delta e_g > 0$)[32]. However, the observed peak energy of $E//c$ for both $e_g$ and $t_{2g}$ is slightly higher than that of $E\perp c$ (*i.e.*, $\Delta t_{2g} < 0$ and $\Delta e_g < 0$), suggesting that the Ti$^{4+}$ undergoes a shift along the $c$-axis away from the TiO$_6$ cluster center. To quantitatively analyze the off-center polar displacement ($\Delta Z_{Ti}$), we performed configurational cluster (CI) calculations[33,34]. When Ti$^{4+}$ is at the center of the TiO$_6$ cluster ($\Delta Z_{Ti} = 0$), the simulated XAS for $E\perp c$ shows higher energy for both $e_g$ and $t_{2g}$ peaks compared to $E//c$ (as shown in the inset of Figure 4B). However, this result doesn't match our experimental XAS findings. To address this, we performed additional calculations by shifting the Ti ion along the $c$-axis, which raises the energy of the $d_{xz}/d_{yz}/d_{3z^2-r^2}$ orbitals. Based on the simulated XAS for different $\Delta Z_{Ti}$ values (Figure S10A), we plotted the changes in simulated $\Delta t_{2g}$ (green line) and $\Delta e_g$ (red line) with $\Delta Z_{Ti}$ (Figure 4B). We clearly observed that the experimental $\Delta t_{2g}$ and $\Delta e_g$ match the simulated green and red lines, respectively, when $\Delta Z_{Ti}$ is around 0.21 Å. Therefore, the best fit for the pure BTO film is achieved at a $\Delta Z_{Ti}$ of about 0.21 Å.

We then redirect our focus to those BFO-BTO solid-solution films with $c/a$ ratio of ~ 1.1. As shown in inset of Figure 4B and Figure S10B, the simulated XAS with Ti$^{4+}$ residing at the TiO$_6$ center shows a giant linear dichroism with peak energy in $E\perp c$ much higher than that in $E//c$, contradicting the experimental XAS results. However, moving Ti$^{4+}$ along the $c$-axis decreases the energy shift from $E//c$ to $E\perp c$ and reverses to from $E\perp c$ to $E//c$ when $\Delta Z_{Ti}$ reaches 0.425 Å. Analyzing the variation of simulated and experimental $\Delta t_{2g}$ and $\Delta e_g$ as the function of $\Delta Z_{Ti}$, as presented in Figure. 4B, $\Delta Z_{Ti}$ for $x$ values between 0.3 and 0.5 is ~0.43 Å. As $x$ further increases to 0.7, $\Delta Z_{Ti}$ decreases to ~0.39 Å. These evaluation reveals that the Ti$^{4+}$ off-center displacements in the solid-solution films are much larger than that in pure BTO film.

Turning our attention to the Fe $L_{3,2}$-edges XAS in Figure 4C, the pure BFO film (with $c/a$ ~1.04) exhibits a slight energy shift from $E//c$ to $E\perp c$. This indicates an elevation of the energy levels of the $d_{xy}/d_{x^2-y^2}$ orbitals and a reduction in the energy levels of $d_{yz}/d_{zx}/d_{3z^2-r^2}$ orbitals in BFO film, which is in accord with strain-induced elongation of the local FeO$_6$ clusters along the $c$-axis. CI calculations are also performed to further rationalize this behavior by shifting Fe$^{3+}$ along the $c$-axis away from the center of the FeO$_6$ cluster[33]. Simulated XAS (Figure 4D and Figure S10C) shows that $Fe^{3+}$ located at the center of $FeO_6$ cluster results in higher peak energy of $E\perp c$ for both $e_g$ and $t_{2g}$ than that of $E//c$. As $Fe^{3+}$ off-center displacement ($\Delta Z_{Fe}$) increases, the energy shift from $E//c$ to $E\perp c$ would decrease. We further depicted the variation of simulated $\Delta t_{2g}$ (green line) and $\Delta e_g$ (red line) with $\Delta Z_{Fe}$ (Figure 4D) according to the simulated XAS (Figure S10C). Notably, experimental $\Delta t_{2g}$ and $\Delta e_g$ can match the simulated green and red lines, respectively, when $\Delta Z_{Fe}$ is ~0.18 Å, very close to the experimental value in bulk BFO (~ 0.17 Å)[35]. For BFO-BTO solid-solution films, one can see in Figure. 4C that the $\Delta e_g$ at Fe $L_{3,2}$-edges increase as $x$ increased. Unlike pure BFO, the solid-solution films experience a significant distortion, which is supposed to further elevate the energy of the $d_{xy}/d_{x^2-y^2}$ orbital and decrease the energy levels of $d_{yz}/d_{zx}/d_{3z^2-r^2}$ orbitals. Consequently, $\Delta t_{2g}$ and $\Delta e_g$ of the solid-solution films are anticipated to become more pronounced compared to pure BFO. However, while a substantial increase in $\Delta e_g$ is observed in comparison to pure BFO film, $\Delta t_{2g}$ remains relatively subtle within the range of $x = 0.3\text{-}0.7$ (Figure 4C-D), suggesting an additional $Fe^{3+}$ off-center displacement. To investigate this phenomenon, we simulated XAS with a $c/a$ ratio of ~1.1 by displacing $Fe^{3+}$ along the $c$-axis. Analysis of the simulated and experimental $\Delta t_{2g}$ and $\Delta e_g$ variations with $\Delta Z_{Fe}$, as depicted in Figure. 4D and Figure. S10D, suggests a fitted $\Delta Z_{Fe}$ of ~0.37 Å for $x = 0.3\text{-}0.5$. As $x$ increases to 0.7, $\Delta Z_{Fe}$ is reduced to ~0.33 Å.

Our analysis of XAS reveals that not only $Ti^{4+}$ but also $Fe^{3+}$ exhibit a more significant off-center displacement in the solid-solution films than in pure BFO and BTO films. This enhanced off-center polar displacement contributes to the polar distortion and increases the ferroelectricity observed in the solid-solution films.

**First-principles calculations**

First principles calculations are performed to further reveal the microscopic origin of the large polar distortion. Figure. 5A shows projected density of state of Bi $6s$ lone pair orbitals in $x = 0.5$ and pure BFO films. One can see that the energy of Bi $6s$ in the highly-strained solid-solution film is blue shifted by 0.57 eV, as compared to pure BFO. As reported in previous literature [36], this blue shift would lead to the enhancement of hybridization between Bi and O ions in the solid-solution film than that in pure BFO film (Figure 5A and Figure S11A). Furthermore, lobe-like charge denstisy near Bi ions implies a lone pair state [36], which shows a stronger lobe-like hybridization in the solid-solution film than pure BFO film. Interestingly, it is clearly observed that charge density between Ti/Fe and O is also enhanced in the solid-solution film compared to that in BTO and BFO films, suggesting enhanced hybridization between Ti/Fe and O ions (Figure 5B and Figure S11-12). The bonding character between Ti/Fe and apical O below the Fermi energy in the solid solution (1-$x$)BFO-$x$BTO film is higher than that in the BTO/BFO film, which is further confirmed by the crystal orbital Hamilton population (COHP) analysis (Figure S13). Similar to other Bi-layered compounds (*e.g.*, $Bi_4Ti_3O_{12}$ and $SrBi_2Ta_2O_9$)[37,38], the orbital hybridization of Ti/Fe and O ions would be enhanced by the greater covalence of Bi-O due to the charge transferring from O to Bi, which induces larger $Ti^{4+}$ and $Fe^{3+}$ displacement in the solid-solution film than that in BTO and BFO films, respectively. Therefore, these results indicate the strained solid-solution films can induce stronger ferroelectricity than BFO and BTO films by enhancing lone pair state characterstics and hybridization between Ti/Fe and O ions in the films.

**DISCUSSION**

Our work demonstrates a strong deviation from Vegard's law and showcases enhanced multiferroic properties in lead-free BFO-BTO solid-solution films. Through a synergistic combination of chemical doping and strain engineering, these solid-solution films exhibit a large tetragonality of ~1.1, along with strong ferroelectric polarization of up to 107 $\mu C/cm^2$ and high Curie temperature of up to 880 °C at $x = 0.2\text{-}0.7$, significantly surpassing those of the individual end members and corresponding bulk materials. Our investigations demonstrate that the strain and chemical mixing-induced enhanced off-center displacement of $Ti^{4+}$ and $Fe^{3+}$ are responsible for the enhanced ferroelectricity. In addition, compared to BFO film, these solid-solution films exhibit enhanced magnetization and an extremely low leakage current density, which is up to four orders-of magnitude lower than that in pure BFO films. These results further demonstrate a great prospect of the strained solid-solution films in lead-free room-temperature multiferroic materials. Our work promises to redefine the landscape of advanced materials and propel us toward innovative frontiers in multiferroic technology.

# EXPERIMENTAL PROCEDURES
## Resource availability
*Lead contact*
Further information and requests for resources should be directed to and will be fulfilled by the lead contact, Zuhuang Chen (zuhuang@hit.edu.cn)

*Material availability*
This study did not generate new unique reagents

*Data and code availability*
All experimental data are available upon reasonable request to the lead contact.

## Epitaxial thin-film growth
(1-$x$)BFO-$x$BTO/SRO heterostructures were grown on (001)-oriented STO single crystal substrates using pulsed laser deposition technique with (1-$x$)Bi$_{1.1}$FeO$_3$-$x$BaTiO$_3$ ceramic targets ($x$ = 0, 0.1, … , 0.8, 1). Note that 10 mol% excess Bi are added into the ceramic targets to compensate for the volatilization of Bi$_2$O$_3$ during high-temperature film growth and target sintering. The ceramic targets were prepared by a high-temperature solid-state reaction route[4,12]. The growth of bottom electrode SRO was carried out at a dynamic oxygen pressure of 22 Pa at 700 °C with a laser fluence of 1 J/cm$^2$ and a laser repetition rate of 5 Hz. The growth of BTO was accomplished at a dynamic oxygen pressure of 2.7 Pa at 620 °C with a laser fluence of 1.1 J/cm$^2$ and a laser repetition rate of 5 Hz. The growth of BFO was performed at a dynamic oxygen pressure of 13 Pa at 700 °C with a laser repetition of 10 Hz and a laser fluence of 0.9 J/cm$^2$. The growth of (1-$x$)BFO-$x$BTO ($x$ = 0.1-0.8) thin films were carried out at a dynamic oxygen pressure of 1 Pa at 640 °C with a laser repetition rate of 5 Hz and a laser fluence of 0.8 J/cm$^2$. After the growth, the samples were cooled to room temperature at 10 °C/min in an oxygen pressure of 75 Torr.

## X-ray diffraction characterizations and properties measurements
X-ray diffraction $2\theta$-$\theta$ scans and reciprocal space mapping studies were obtained by a Rigaku smartlab diffractometer using Cu $K_\alpha$ radiation. For electric properties measurements, 50 nm thick Pt top electrode with a diameter of 30 μm was deposited on the surface of thin films using a magnetron sputtering equipment (Arrayed Materials RS-M). To accurately determine ferroelectric polarization values of (1-$x$)BFO-$x$BTO thin films, Pt electrode area was calibrated accurately using scanning electron microscopy (SEM, Hitachi SU8010) and the result is exhibited in Figure S14. The polarization-electric field hysteresis loops were measured using aixACCT TF 3000. The leakage behavior was measured using a keithley 4200A-SCS semiconductor parameter analyzer. The optical band gap of thin films is studied using the UV-VIS-NIR spectrophotometer (Simadzu, UV-3600) over the wavelength range 300-800 nm. In order to obtain the magnetic properties of the (1-$x$)BFO-$x$BTO films, the magnetic field was applied parallel to the sample plane using a quantum design superconducting quantum interference device (SQUID).

## STEM characterizations
Cross-sectional specimens intended for TEM and STEM analysis were fabricated through a series of processes including slicing, grinding, and dimpling, followed by final ion milling utilizing the Gatan Precision Ion Polishing Systems 695. The iDPC STEM images were obtained by ThermoFihser Spectra 300 micoscope equiped with a high-brightness field-emission gun and double aberration correctors operated in iDPC-STEM mode with a probe convergence semiangle of 25 mrad. The images were captured by a segmented iDPC STEM detector with a collection angle ranging from 7 to 27 mrad. EELS spectrum imaging (SI) data across the (1-$x$)BFO-$x$BTO/SRO interface were collected over a range of 150 eV to 3200 eV with a dispersion of 0.9 eV/channel, utilizing a Gatan K3 camera within a Gatan Continuum GIF system to acquire comprehensive EELS information for Ti, O, Fe, Ba, Ru, Sr, and Bi elements. An entrance aperture of 5 mm was employed for the measurements.

**X-ray absorption spectroscopy**
X-ray absorption spectroscopy (XAS) was performed in the total electron yield (TEY) mode at the TLS11A and TPS45A beamlines of the National Synchrotron Radiation Research Center (NSRRC) in Taiwan. X-ray linear dichroism (XLD) measurements were obtained by studying the difference between horizontal and vertical linear polarized XAS using the TEY mode. The incident X-ray beam was angled at 20° with respect to the sample surface. All measurements were taken at room temperature. During the Fe and Ti $L_{3,2}$ XAS measurements, spectra of $Fe_2O_3$ and $SrTiO_3$ were also measured simultaneously to ensure proper energy calibration.

**Full multiplet configurational interaction (CI) cluster calculation**
This theoretical approach considers the fully atomic multiplet effect and configurational interaction as well as local solid effect. It is a well-proven and well-established approach for simulating XAS spectra [33,34,39,40]. Three configurations, namely $|d^n\rangle, |d^{n+1}\underline{L}^1\rangle, |d^{n+2}\underline{L}^2\rangle$ with n = 0 for $Ti^{4+}$ and n = 5 for $Fe^{3+}$ are employed as the ground state calculations, where $|\underline{L}^n\rangle$ denotes the number of the holes on the ligand orbitals that make the bond with the $Ti^{4+}/Fe^{3+}$ ion. These ligand orbitals are formed by the linear combination of oxygen 2p orbitals with respect to the local symmetry of the $FeO_6$ and $TiO_6$ cluster. Three additional configurations, namely $|\underline{c}d^{n+1}\rangle, |\underline{c}d^{n+2}\underline{L}^1\rangle, |\underline{c}d^{n+3}\underline{L}^2\rangle$ with n = 0 for $Ti^{4+}$ and n = 5 for $Fe^{3+}$ are further employed as the final state to calculate the XAS, where $\underline{c}$ denotes a hole that appeared in the $Fe^{3+}/Ti^{4+}$ 2p core level after an electron transition to the 3d orbital by X-ray. We define the energy center of each configuration as $|d^n\rangle = 0, |d^{n+1}\underline{L}^1\rangle = \Delta$, $|d^{n+2}\underline{L}^2\rangle = 2\Delta + U_{dd}$, and $|\underline{c}3d^n\rangle = \varepsilon_{XAS}, |\underline{c}3d^{n+1}\underline{L}^1\rangle = \varepsilon_{XAS} + \Delta + U_{dd} - U_{dp}, |\underline{c}3d^9\underline{L}^2\rangle = \varepsilon_{XAS} + 2\Delta + 3U_{dd} - 2U_{dp}$, where the charge transfer energy Δ = 0.5/2.0 eV for $Ti^{4+}/Fe^{3+}$, the monopole Coulomb repulsion $U_{dd}$ = 5.0 eV for the both $Ti^{4+}$ and $Fe^{3+}$, and $U_{pd}$ = 5.5/6.0 eV for $Ti^{4+}/Fe^{3+}$ were used in the calculations. The Slater integrals for accounting multiplet splittings of the Coulomb interactions were determined with the use of R. D. Cowans coded RCN36K[41]. The $\varepsilon_{XAS}$ is the energy required to excite an electron from the 2p to the 3d orbital. The coefficient of d-d and d-p spin-orbital coupling were determined with the use of R. D. Cowans coded RCN36K [41]. The Madelung potential is employed to calculate the 3d crystal field splitting of $Ti^{4+}$ and $Fe^{3+}$ ions, treating the near ligand atoms as the point charge. The hybridization strength between the Ti/Fe 3d and O 2p orbitals was computed for different Ti/Fe–O bond lengths following Harrison's framework[42].

**First-principles calculations**
The first-principles calculations have been performed by using the plane wave basis Vienna ab initio simulation package (VASP) based on density functional theory (DFT) [43]. The Perdew–Burke–Ernzerhof (PBE) exchange-correlation functional was employed for the generalized gradient approximation (GGA) [44]. The ion-electron interaction was treated by the projector augmented wave (PAW) method[45]. The Bi $5d^{10}6s^26p^3$, Ba $5s^25p^66s^2$, Fe $3p^63d^64s^2$, Ti $3p^63d^24s^2$ and O $2s^22p^4$ are treated as valence electrons. The cutoff energy of plane-wave basis was set to 550 eV, and a 6 × 6 × 6 Γ-centered k-point grid was adopted for the k-point sampling for the 2 × 2 × 2 (40-atom) supercell. All structures are fully relaxed until the force on each atom was less than 1 meV/Å and the total energy difference between the two adjacent steps was less than $10^{-7}$ eV. To treat localized d orbitals, we applied Hubbard U corrections with $U_{eff}$ = 4 eV for Fe atoms, which has been performed in previous research [46-48]. The in-plane lattice constants of (1-x)BFO-xBTO films are fixed to the in-plane lattice constant of STO, to mimic the experiments of thin film on STO substrate.

**SUPPLEMENTAL INFORMATION**
Supplemental information can be found online

**ACKNOWLEDGMENTS**
This work was supported by National Natural Science Foundation of China (Grant No. 52372105), Shenzhen Science and Technology Innovation project (Grant No. JCYJ20200109112829287), and Shenzhen Science and Technology Program (Grant No. KQTD20200820113045083). Z.H.C. has been supported by State Key Laboratory of Precision Welding & Joining of Materials and Structures (Grant No. 24-Z-13), "the Fundamental Research Funds for the Central Universities" (Grant No. HIT.OCEF.2022038) and "Talent Recruitment Project of Guangdong" (Grant No. 2019QN01C202). M. J. Z. acknowledges the Guangdong Basic and Applied Basic Research Foundation (2022A1515110086), C.-Y.K. acknowledges the financial support from the Ministry of Science and Technology in Taiwan under grant no. MOST 110-2112-M-A49-002-MY3. C.F.C. and C.-Y.K. acknowledge support from the Max Planck-POSTECH-Hsinchu Center for Complex Phase Materials. S.D. acknowledges Science and Engineering Research Board (SRG/2022/000058; EEQ/2023/000089), and Indian Institute of Science start up grant for financial support. W.L. and L.B. thank the Vannevar Bush Faculty Fellowship (VBFF) Grant No. N00014-20-1-2834 from the Department of Defense, and is thankful for support from the MonArk Quantum Foundry supported by the National Science Foundation Q-AMASE-i program under NSF Award No. DMR-1906383.

**AUTHOR CONTRIBUTIONS**
Z.H.C. conceived and supervised this study; T.W. fabricated the films and performed the XRD measurements with the assistance of Y.W.Z. and S.P.; T.W. carried out electric and magnetic properties measurements with the assistance of H.L.H. and Z.D.X.; M.J.Z., Y.L.T. and X.L.M. performed the STEM characterizations; C.-Y.K. and C.-F.C. conducted the XAS measurements; D.H.Z., W.L., Y.R.Y. and L.B. performed first-principle calculations; S.D., L.C. and X.J.L. provided insights and interpretation of the results; T.W., C.-Y.K., Y.R.Y. and Z.H.C. wrote the manuscript; All authors discussed the data and commented on the manuscript.

**DECLARATION OF INTERESTS**
The authors declare no conflict of interest.

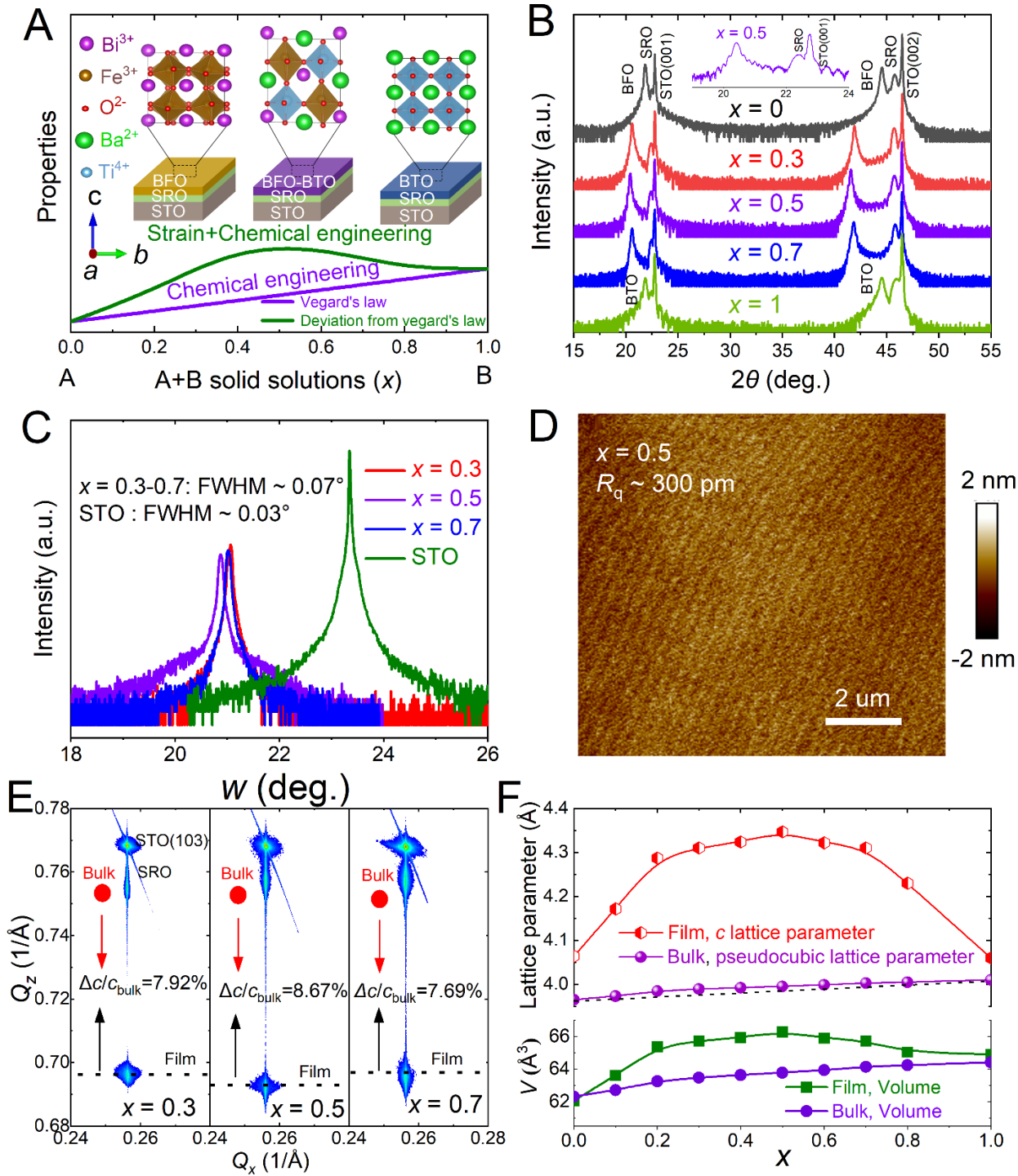

**Figure 1. Enhanced tetragonality and deviation of Vegard's law in (1-$x$)BFO-$x$BTO thin films.**
**A,** Schematic diagram illustrating the varization in properties of a solid solution as a function composition $x$. The inset shows enhanced tetragonality in BFO-BTO multiferroic films by strain and chemical engineering. **B,** Composition-dependent XRD $\theta$-$2\theta$ scans of BFO-BTO thin films. **C,** The $\omega$-scans (rocking curves) around the 002-diffraction condition of both the films and substrate. **D,** AFM topographic image of the BFO-BTO solid-solution films with $x = 0.5$. **E,** Reciprocal space mappings studies of the BFO-BTO films about the 103- diffraction conditions. **F,** Composition-dependent variations in the $c$ lattice parameter and unit-cell volume of BFO-BTO films, as well as the bulk pseudocubic lattice parameter and unit-cell volume.

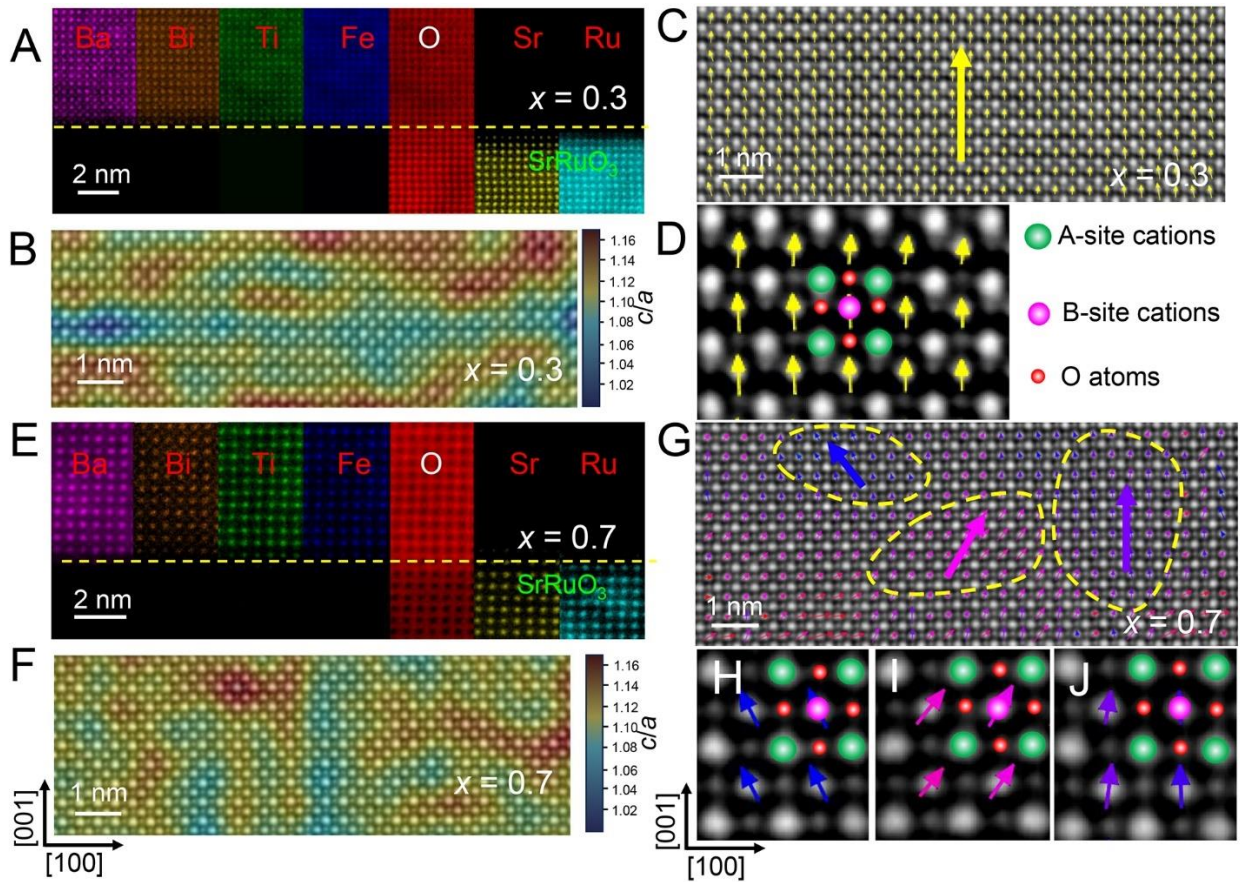

**Figure 2**. Compositional and structural characterizations using atomic-resolution STEM. **A** and **E**, Atomic EELS chemical mapping of Ba- and Ru-*M* edges, Bi-, Sr-, Ti- and Fe-*L* edges, O-*K* edge at the interfaces the BFO-BTO/SRO heterostructures. **B** and **F,** *c*/*a* ratio maps overlaid with the iDPC-STEM images; **C** and **G**, Superposition of the displacement vectors (yellow arrows) of B-site atoms relative to the center of four neighboring A-site cations with the experimental iDPC images of the films; **D** and **H-J**, magnified view of local polar bevaviors in **C** and **G**, respectively. **A-D**: $x = 0.3$; **E-J**: $x = 0.7$.

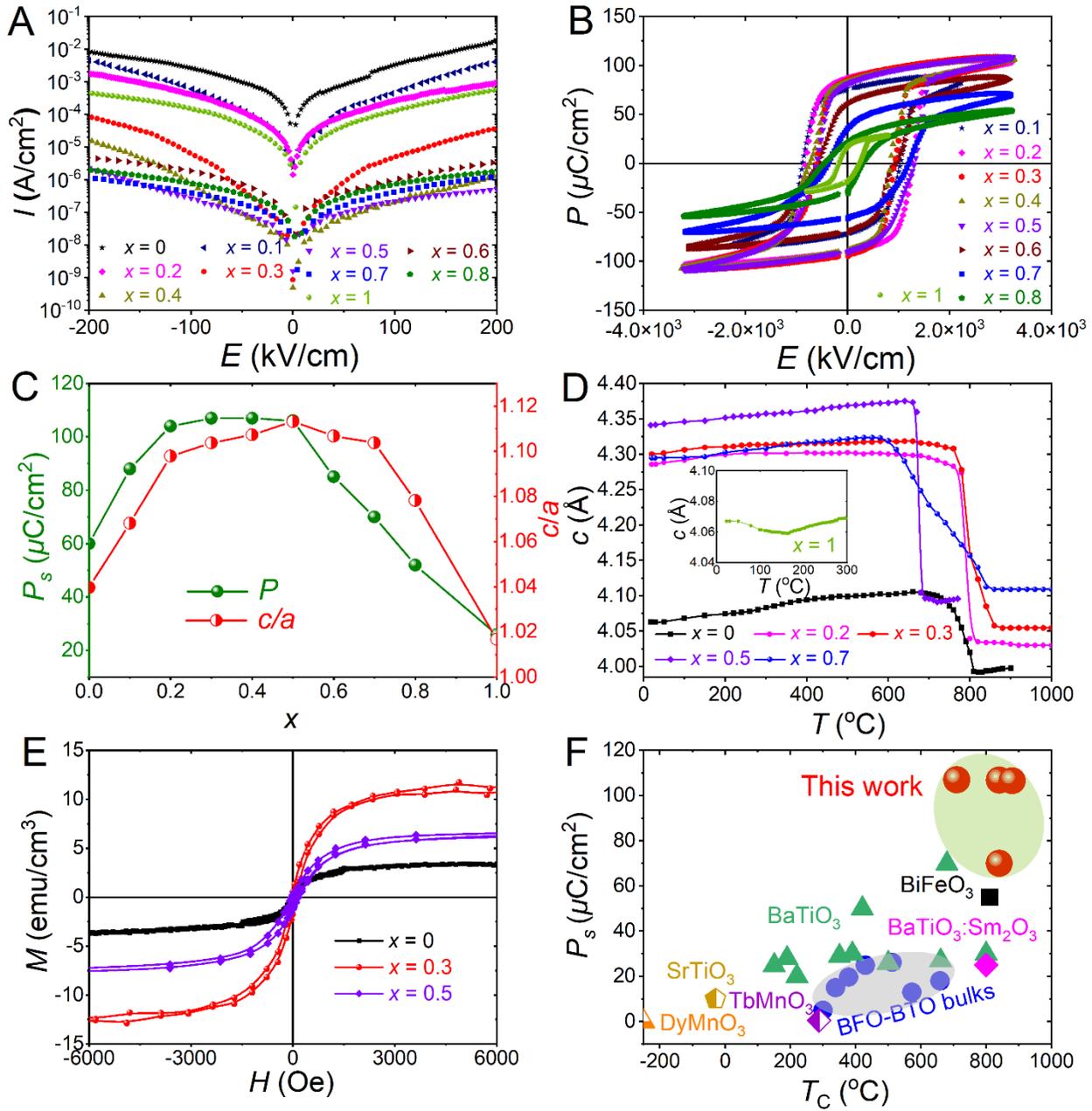

**Figure 3**. **Enhanced properties of the strained (1-$x$)BFO-$x$BTO solid-solution films.** **A**, Leakage current as a function of applied electric field for BFO-BTO thin films; **B**, Polarization-electric field hysteresis loops for BFO-BTO thin films at 50 kHz and room temperature. **C**, The variation of $P_S$ and $c/a$ with $x$. The $P_S$ value of BFO film is obtained from thick film (Figure S8B); **D**, Temperature-dependence of out-of-plane $c$ lattice parameter for the BFO-BTO thin films; **E**, Magnetic hysteresis loops of (1-$x$)BFO-$x$BTO films at 300 K; **F**, The $T_C$ and $P_S$ comparison of BFO-BTO thin films with corresponding bulks and other representative (001)-oriented lead-free perovskite multiferroic/ferroelectric films. Materials include BFO-BTO bulks [4,13] and thin films: BiFeO$_3$, BaTiO$_3$[49-51], BaTiO$_3$:Sm$_2$O$_3$[52], TbMnO$_3$ [53], DyMnO$_3$[54], SrTiO$_3$[55].

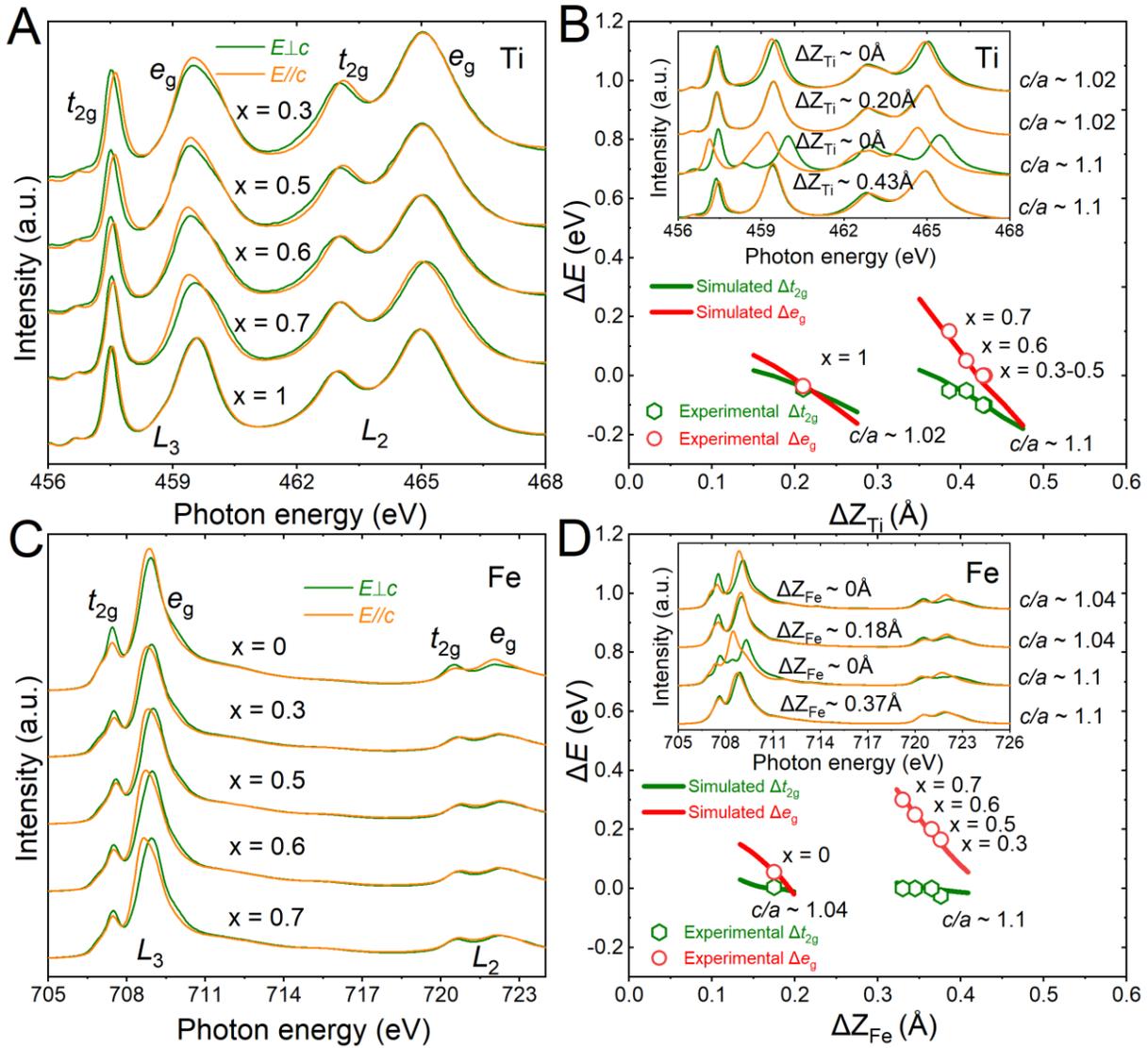

**Figure 4**. **Linearly polarized X-ray absorption spectroscopy studies of atomic displacement.** **A**, Experimental Ti $L_{3,2}$-edges XAS of BFO-BTO films. **B**, The variation of $\Delta t_{2g}$ and $\Delta e_g$ with $\Delta Z_{Ti}$. The inset shows simulated Ti $L_{3,2}$-edges XAS of BFO-BTO films. **C**, Experimental Fe $L_{3,2}$-edges XAS of BFO-BTO films. **D**, The variation of $\Delta t_{2g}$ and $\Delta e_g$ with $\Delta Z_{Fe}$. The inset show simulated Fe $L_{3,2}$-edges XAS of BFO-BTO films. The simulated XAS spectra is performed by employing $c/a$ based on XRD results. Green line: the variation of simulated $\Delta t_{2g}$ with ionic displacement; Red line: the variation of simulated $\Delta e_g$ with ionic displacement; Green hexagon: experimental $\Delta t_{2g}$; Red circular: experimental $\Delta e_g$.

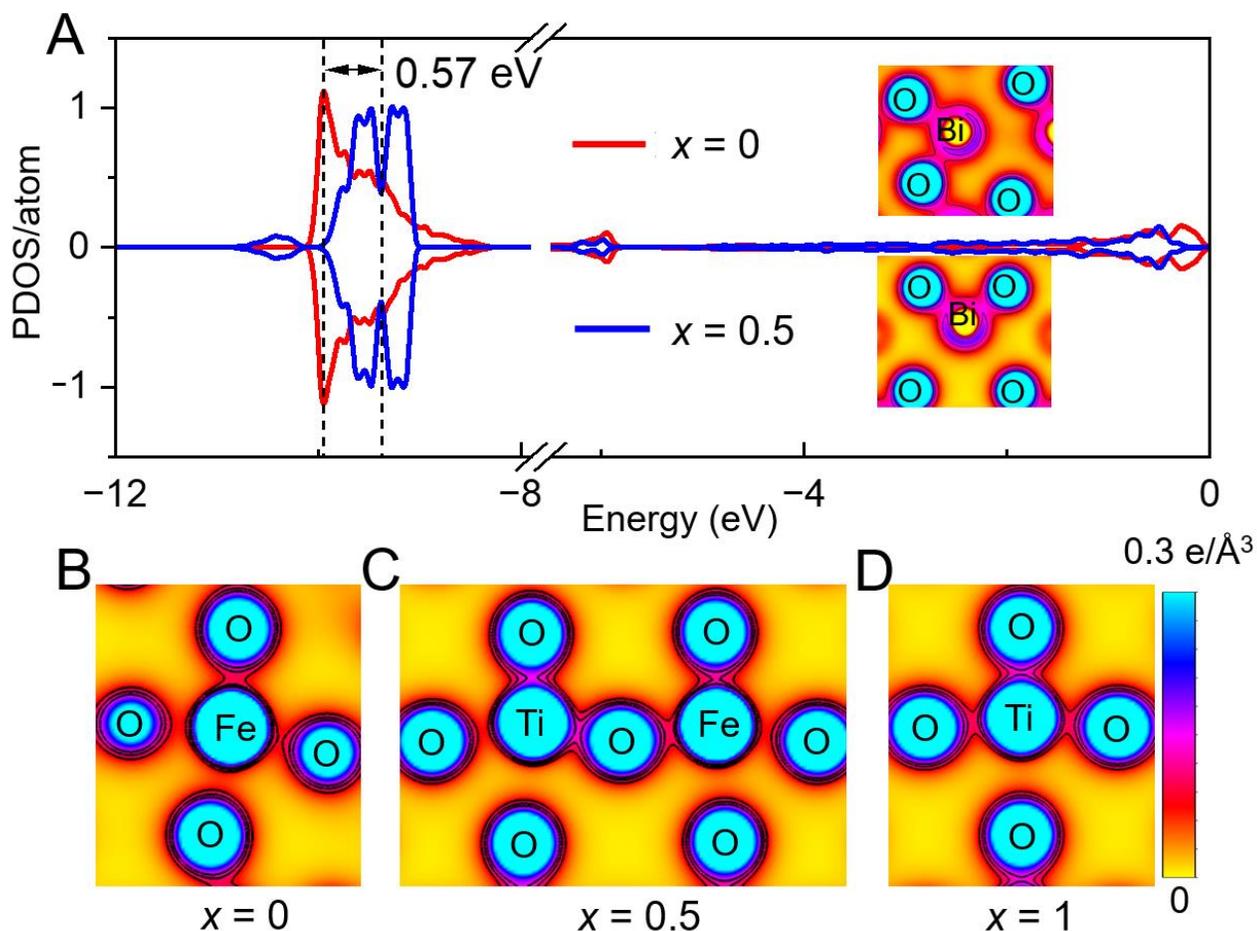

**Figure 5. The first-principles calculations of (1-$x$)BFO-$x$BTO films. A,** The projected density of state (PDOS) of Bi 6$s$ orbital in BFO ($x$ = 0) and (1-$x$)BFO-$x$BTO ($x$ = 0.5) films for states in the energy range from -12 eV to 0 eV. Note that the Fermi level at valence band maximum is set as the origin of the energy axis (0 eV). The upper (lower) inset shows the total charge density around the Bi ions for BFO (0.5BFO-0.5BTO) film. The lobe-like isosurface near Bi ion indicates the lone pair state. The film with $x$ = 0.5 exhibits an enhanced lone pair hybridization compared to that in BFO films. **B**, **C** and **D** show the total valence electron density distributions around the Fe (Ti) atoms on the $ac$ plane for $x$ = 0, 0.5, 1, respectively. The film with $x$ = 0.5 exhibits stronger hybridization between Fe 3$d$ (Ti 3$d$) and O 2$p$ than that in BFO (BTO) films.